\documentclass[useAMS,usenatbib]{mn2e}

 \usepackage{epsfig}
 \usepackage{hyperref}
 \usepackage{ifpdf}
\usepackage{amsmath}
\usepackage{amssymb}
 \usepackage{adjustbox}
 \usepackage{blindtext}

\usepackage{textcomp}

 \usepackage{adjustbox}
 \usepackage{blindtext}
\usepackage{lmodern}
\usepackage{slantsc}

\newcommand{\aap}{A\&A}          
\newcommand{\aj}{AJ}             
\newcommand{\apj}{ApJ}           
\newcommand{\apjs}{ApJS}         
\newcommand{\apjl}{ApJL}         
\newcommand{\mnras}{MNRAS}       
\newcommand{\aaps}{A\&AS}        
\newcommand{\pasp}{PASP}         
\newcommand{\araa}{ARAA}         

\title[Physical regimes for giant H{\textsl{\textsc{ii}}} regions and GMCs in the Antennae]
{Two physical regimes for the giant H{\sc{ii}} regions and giant molecular clouds in the Antennae Galaxies}
\author[Zaragoza-Cardiel et al.]{J. Zaragoza-Cardiel$^{1,2}$\thanks{E-mail:jzc@iac.es},
 J. Font$^{1,2}$, J. E. Beckman$^{1,2,3}$,
 B. Garc\'ia-Lorenzo$^{1,2}$,\newauthor
 S. Erroz-Ferrer$^{1,2}$, and L. Guti\'errez$^{4}$\\
$^{1}$Instituto de Astrof\'isica de Canarias, C/ V\'ia L\'actea s/n, 38205 La Laguna, Tenerife, Spain\\
$^{2}$Department of Astrophysics, University of La Laguna, E-38200 La Laguna, Tenerife, Spain\\
$^{3}$CSIC, 28006 Madrid, Spain\\
$^{4}$Universidad Nacional Aut\'onoma de M\'exico, Instituto de Astronom\'ia, Ensenada, B. C., Mexico}

\begin{document}

\label{firstpage}

\maketitle

\begin{abstract}

We have combined observations of the Antennae galaxies from the radio interferometer ALMA 
(Atacama Large Millimeter/submillimeter Array) and from the optical interferometer 
GH$\alpha$FaS (Galaxy H$\alpha$ Fabry-Perot System). 
The two sets of observations have comparable angular and spectral resolutions, 
enabling us to identify 142 giant molecular clouds (GMCs) and 303 H{\sc{ii}} regions. 
We have measured, and compared, their basic physical properties (radius, 
velocity dispersion, luminosity). For the H{\sc{ii}} regions, we find two physical 
regimes, one for masses $>10^{5.4} \mathrm{M_{\odot}}$ of ionized gas,  where the gas density 
increases with gas mass, the other for masses $<10^{5.4} \mathrm{M_{\odot}}$ of ionized gas, where 
the gas density decreases with gas mass. For the GMCs, we find, in contrast 
to previous studies in other galaxies over a generally lower mass range of clouds, 
that the gas surface density increases with the radius, hinting at two regimes 
for these clouds if we consider both sources of data. We also find that the 
GMC mass function has a break at $10^{6.7}\mathrm{M_{\odot}}$. Using the velocity dispersion 
measurements, we claim that the difference between the regimes is the nature 
of the dominant binding force. For the regions in the lower mass range, the 
dominant force is the external pressure, while in the higher mass range it 
is the internal gravity of the clouds. In the regime where gravity is dominant,  
the star formation rate, derived from the dust-corrected H$\alpha$ luminosity, 
increases super-linearly with the velocity dispersion, and the gas density 
increases with the gas mass.
\end{abstract}


\section{Introduction}

Star formation takes place largely if not exclusively within molecular clouds
\citep{2011EAS....51..133K,2012arXiv1210.6990S}. Studies of the
statistical properties of molecular clouds
help us to understand the processes involved in star formation, in both Galactic
\citep{1981MNRAS.194..809L,2009ApJ...699.1092H,2010ApJ...723..492R,2010ApJ...716..433K}
and extragalactic \citep{2008ApJ...686..948B,2012ApJ...750..136W,2014ApJ...784....3C} 
environments. 

When the first massive stars form, the gas is heated and ionized leading 
to the formation of H{\sc{ii}} regions. This may cause the quenching of star formation via local 
gas stripping. However, if the original molecular clouds were sufficiently massive, 
and thus strongly bound, the effects of this heating in disrupting the original cloud   
could be relatively moderate \citep{2012MNRAS.424..377D}.

In addition, the detailed effects of galaxy interactions on star formation
demand considerable further research \citep{2011EAS....51..107B}. Studying the properties 
of molecular clouds in galaxy interactions can reveal important clues about star formation 
and its role in the evolution of galaxies.

\cite{1981MNRAS.194..809L} studied the properties (luminosity/mass, 
size, and velocity dispersion) of 54 molecular clouds in the Galaxy, and found 
correlations between them. Nowadays, we call those correlations  Larson's laws, 
and we still do not know if they are universal. 
The correlation known as the third law 
is the mass size relation, which Larson found to follow $M\propto R^{1.9}$, i. e., 
the molecular gas surface density is approximately constant with size. The exponent 
in this relation has varied from study to study but always around 2 and considerably
less than 3 
\citep{2010ApJ...723..492R,2010A&A...519L...7L,2010ApJ...716..433K,2014ApJ...784....3C}. 
 \cite{2014ApJ...784....3C} found 
that the properties of the M51 giant molecular clouds (GMCs) apparently 
differ in the arms, the interarm regions and the circumnuclear zone. 
Statistical studies of molecular clouds can help us to study the star formation behaviour 
in different environments and on different scales.  The study of the properties of GMCs in
interacting galaxies is harder than in isolated galaxies due to the lack of suitable nearby examples. M51 is 
the closest pair of galaxies, however, the Antennae galaxies are the closest gas-rich major merger 
pair of galaxies. Thus, 
improving the spatial resolution and the sensitivity of observations of the Antennae could 
reveal new properties of the gas clouds.

Statistical studies of H{\sc{ii}} regions have been performed, but never compared explicitly with the equivalent 
properties of molecular clouds. 
 There are, however, correlations between parameters (luminosity, 
size, and velocity dispersion) of H{\sc{ii}} regions which are equivalent to those for molecular clouds. 
The exponent $Q$ in the H$\alpha$ luminosity-size relation, $L_{{\mathrm H\alpha}}\propto R^{Q}$, has  usually 
been found to have values of  around 2 and significantly less than 3 \citep{1981MNRAS.195..839T,2011AJ....141..113G}. 
The luminosity-velocity dispersion relation has been proposed as a useful distance 
calibrator \citep{1987MNRAS.226..849M, 2014arXiv1405.4010C}, but we still do not know  
what causes the width of the H{\sc{ii}} region velocity dispersions to change. Recent results are contradictory, pointing  
towards self-gravity \citep{2013MNRAS.432..998Z,2014arXiv1405.4010C} or 
towards the energy injected into the interstellar medium by the ongoing star formation
\citep{2014arXiv1405.5731M}.

The break in the luminosity function 
\citep{1989ApJ...337..761K,2006A&A...459L..13B} at 38.6 dex 
has also been proposed as a distance calibrator. Defining the physical basis of the 
luminosity-velocity dispersion relation and the 
luminosity function break would clearly be important to validate them as distance calibrators.

The Antennae consist of a pair of galaxies in obvious interaction, which is classified as a luminous system in the infrared 
(LIRG, $L_{IR}=10^{11}$; \cite{2003AJ....126.1607S}). The two are in a fairly 
early stage of a merger process, so it is commonly accepted that, in the future, the system will 
develop into an ultra-luminous infrared galaxy (ULIRG). The luminosity is dominated by the 
infrared luminosity and the Antennae present off-nuclear obscured star forming regions 
\citep{1998A&A...333L...1M,2004ApJS..154..193W}. This makes the Antennae galaxies a 
particularly suitable system for understanding the role of star formation in the
evolution of galaxies, as well as the formation of the off-nuclear peaks of star 
formation in galaxy mergers.

Previous authors have used the high resolution of the Hubble Space Telescope 
(\emph{HST}) data to resolve super star clusters (SSC) in the Antennae 
\citep{2007MNRAS.377...91A,2010AJ....140...75W}, which gave rise to contradictory 
results about the presence of a turnover in the SSC luminosity function.

Recently, \cite{2012ApJ...750..136W} measured the physical parameters of molecular clouds in this
interacting galaxy pair (also referenced as Arp 244)  based on a combination of observations of 
SMA (Submillimeter Array) and PdBI (Plateau de Bure Interferometer), yielding a synthesized angular resolution 
of $3.3\mathrm{arcsec}\times1.5\mathrm{arcsec}$. They found that the mass
function of the clouds  identified in the CO(2-1) line emission shows a double peaked distribution,
 with a break at $\sim10^{6.5}\rm{M_{\odot}}$.
They suggested that the clouds with masses below this value share the properties 
(essentially the dependence of mass and velocity dispersion on cloud radius) 
of the GMC's in normal star forming galaxies. The high
mass cloud population is associated with sites of peak star formation. 
However, \cite{2012ApJ...745...65U} described the molecular cloud population for clouds identified in the CO(3-2) 
line emission, with an angular resolution of $1.42\mathrm{arcsec}\times1.12\mathrm{arcsec}$ based on observations 
of SMA, and they made no comment about finding two molecular cloud populations. 
Recent results by 
\cite{2013MNRAS.432..998Z} showed two populations of H{\sc{ii}} regions in another interacting galaxy 
system Arp 270 which, the authors suggested, are the result of the evolution of 
two populations of molecular clouds. 

In the present paper, we present a comparison of the observed properties of the molecular clouds and 
H{\sc{ii}} regions in Arp 244 inferring what this comparison can tell us about the evolution of 
massive molecular clouds into luminous H{\sc{ii}} regions.
For this purpose, we combined data from the ALMA (Atacama Large Millimeter/submillimeter 
Array) radio interferometer for the molecular clouds, and
the Fabry-Perot optical interferometer GH$\alpha$FaS (Galaxy H$\alpha$ Fabry-Perot system)
for the H{\sc{ii}} regions.

In section \S2 we present ALMA observations and the GH$\alpha$FaS observations and data reduction. 
In section \S3 we show how we have estimated the different properties of the H{\sc{ii}} regions and the GMCs. 
In section \S4 the results relative to the proposed presence of two regimes of star forming regions are shown.
In section \S5 we give evidence supporting a scenario where the more massive regions are
gravitationally bound rather than being bounded by the external pressure. Finally, in
section \S6 we present our conclusions.

\section{Observations}

\begin{figure*}

\centering
\epsfig{file=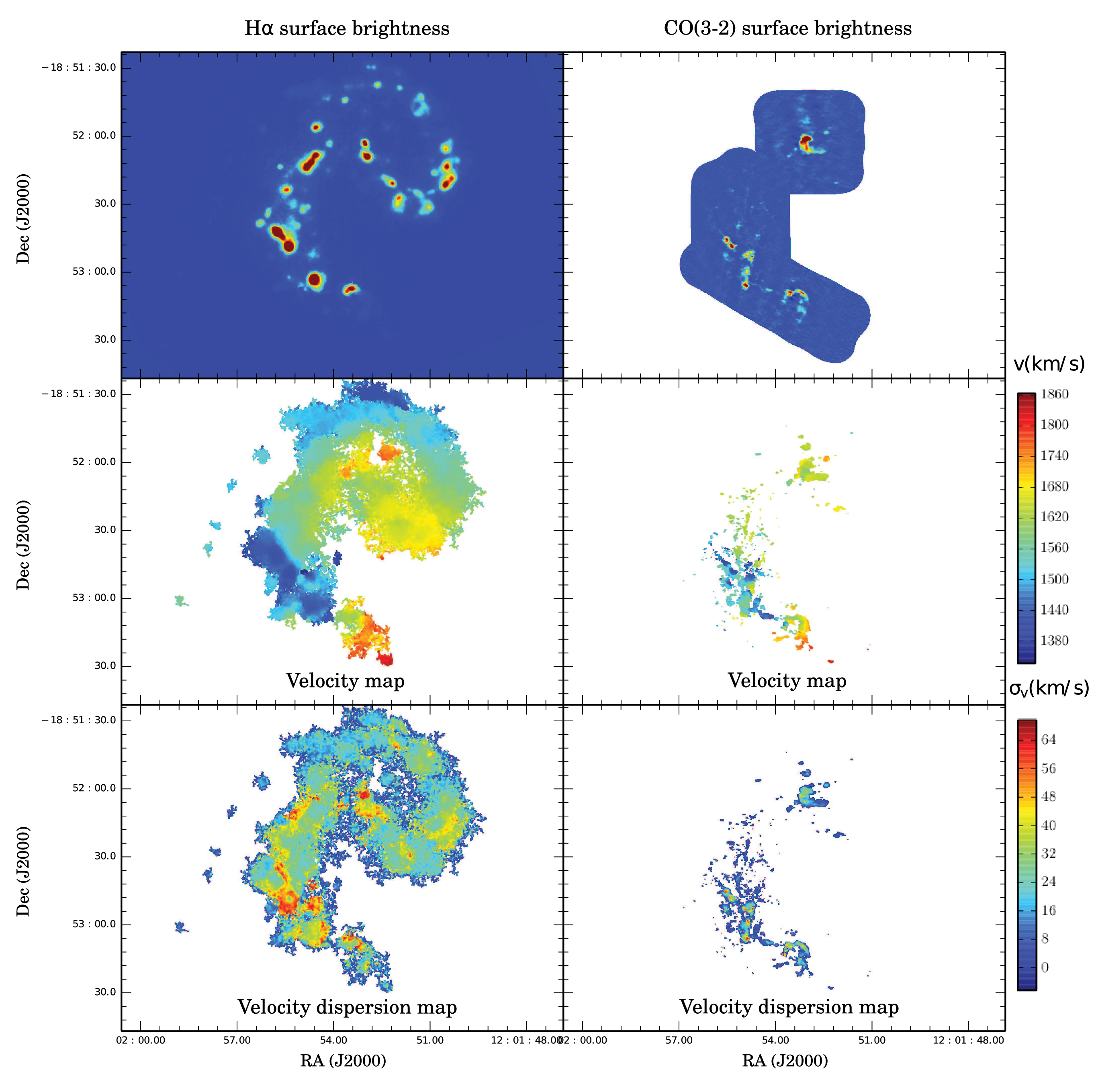,width=0.95\linewidth}
%
 \put(-410,461.5){\large (a)}
%
%
 \put(-230,461.5){\large (b)}

\caption{Moment maps of The Antennae galaxies. (a) Derived from the GH$\alpha$FaS datacube. (b)
Derived from the ALMA datacube. The H$\alpha$ datacube as well as the product moment maps are available through CDS.}
\label{fig_moments}
\end{figure*}

We present two sets of observations of the Antennae: data cubes of emission in H$\alpha$
from the ionized component of the interstellar medium, and of emission in the CO(3-2) line from
the molecular component.

\subsection{ALMA observations}

We used three different mosaics of the Antennae system from ALMA, covering three zones.
The first two mosaics of the observations in the CO(3-2) line are from the ALMA Science Verification
programme\footnote{\hyperref[http://almascience.eso.org/almadata/sciver/AntennaeBand7/]
{\url{http://almascience.eso.org/almadata/sciver/AntennaeBand7/}}}. The third mosaic in the 
CO(3-2) line is from the
ALMA Early Science Cycle 0 archive 
(project ADS/JAO.ALMA\#2011.0.00876.S)\footnote{\hyperref[http://almascience.eso.org/alma-data/archive]
{\url{http://almascience.eso.org/alma-data/archive}}}. 
 The Science Verification data have been described in \cite{2012A&A...538L...9H}. The Early Science mosaic 
is derived from a series of observations from 2012 July to November. The total 
on source time is $3$ h with 14 to 24 antennas and baselines from $21$ to $402$ m. We have 
used the reduced mosaic published in the archive where the raw data have been corrected for 
atmospheric effects, and flux, bandpass, phase, and amplitude calibrations have been applied. We then corrected 
 the Science Verification and the Early Science mosaics for the primary beam response.

The angular resolution for both mosaics is superior 
(synthesized beam $\sim 0.6\rm{arcsec}\times 1.1\rm{arcsec}$
 for the Science Verification data, synthesized beam $\sim 0.4\rm{arcsec}\times 0.7\rm{arcsec}$ 
for 
the Early Science data)  to those of previous observations \citep{2012ApJ...745...65U,2012ApJ...750..136W}. Assuming
a distance to the Antennae of $22\mathrm{Mpc}$ \citep{2008AJ....136.1482S}, the
synthesized beam size $ 64\mathrm{pc}\times117\mathrm{pc}$ for 
the two first mosaics (Science Verification data) and $ 41\mathrm{pc}\times61\mathrm{pc}$ for
the Early Science,
while the velocity is binned in channels
of $10 \rm{km s^{-1}}$ in both cases. 
 Larger than $420$pc structures for Early Science and larger 
than $354$pc structures for Science Verification are not detected due to the lack of short and
zero spacing. However, 
structures of this size or larger are not required for our analysis as we will see below.
The
ALMA maps covered the areas shown in Fig. \ref{fig_moments} (b),  where we show the moment maps derived. 
CO surface brightness is shown as moment 0, (integrated intensity), the velocity map 
as moment 1 (intensity weighted velocity field), and the velocity dispersion map as 
moment 2 (intensity weighted velocity dispersion). The velocity and velocity dispersion maps have been masked 
in order to consider only locations where the line is stronger than a minimum detectable value: $4\times l_{\mathrm{rms}}$.   

 The rms in a single channel is $l_{\mathrm{rms}}=3\thinspace\mathrm{mJy/beam}$ for the 
Science Verification data and $l_{\mathrm{rms}}=2\thinspace\mathrm{mJy/beam}$ for the Early Science data with a channel 
size of $10\mathrm{km/s}$ in both cases. We have used the three mosaics 
separately, and in the region where Early Science and Science Verification data 
overlap, we use Early Science data because of its better sensitivity and resolution. This 
yields an inhomogeneous data set, and this is why we use a method unbiased with respect 
to sensitivity and resolution,  
as described below, to analyse the data. 

\subsection{GH$\alpha$FaS observations}

Our observations
in H$\alpha$ were taken with the GH$\alpha$Fas Fabry-Perot interferometer \citep{2008PASP..120..665H} on
the $4.2\rm{m}$ William Herschel Telescope (WHT) at
the Observatorio del Roque de los Muchachos, La Palma. 
 on the night of 2012 January 24. 
The seeing limited angular resolution 
of $0.9 \rm{arcsec}$,
$96pc$ at the assumed distance of $22\mathrm{Mpc}$, was
very close to that of the ALMA CO observations, while the velocity resolution,
at $8 \rm{km s^{-1}}$ is also similar to that in the ALMA observations. 
 The velocity range of the observation is from 1029 to 2211km/s.

The H$\alpha$ map, shown in Fig. \ref{fig_moments} (a),
covered an area on the sky of $3.4\times3.4 \rm{arcmin}^2$. For
details of the data processing of the Fabry-Perot data we refer the reader to
\cite{2010MNRAS.407.2519B}.

A continuum-subtracted and flux-calibrated ACAM (auxiliary-port camera) H$\alpha$ image was used to
perform the flux calibration of the GH$\alpha$FaS data. ACAM is an instrument
mounted permanently at the WHT used for broad-band and narrow-band imaging.
Fluxes from selected H II regions in both the GH$\alpha$FaS cube and the ACAM image were measured, and 
then compared following the procedure explained in \cite{2012MNRAS.427.2938E}.

 Calibrating in velocity and phase adjustment yields a data cube to which we  
applied the procedures described in \cite{2006MNRAS.368.1016D} to derive maps of 
Arp 244 in H$\alpha$ surface brightness, velocity and velocity dispersion, which are shown
in Fig. \ref{fig_moments} (a). The reduced H$\alpha$ data cube and the derived 
moment maps are available through CDS.  The rms in a single channel is 
$l_{\mathrm{rms}}=8\times10^{-19}\mathrm{erg\thinspace s^{-1}\thinspace cm^{-2}\thinspace \AA{}^{-1}}$ 
($1.1\thinspace\mathrm{\mu Jy}$) with 
a channel size of $8\mathrm{km/s}$.

\section{Properties of H{\sc{ii}} regions and GMC\textsc{s}}

 We used \textsc{dendrogram}\footnote{\href{http://www.dendrograms.org}{http://www.dendrograms.org}}, 
a \textsc{python} package to compute \textgravedbl dendrograms\textacutedbl of astronomical data
\citep{2008ApJ...679.1338R} to derive the luminosity, radius and
velocity dispersion of the H{\sc{ii}} regions and molecular clouds using the GH$\alpha$FaS and
ALMA datacubes, respectively. To facilitate the reading of this paper we describe briefly how the 
properties are extracted from the datacubes. For more details we refer the reader to the original 
sources, \cite{2006PASP..118..590R,2008ApJ...679.1338R}, of which our description below is a summary. 

Both ALMA and GHaFaS datacubes can be described as a collection of values, $T_i$, 
at a given 2D spatial point, ($x_i,y_i$), and 
at a given velocity, $v_i$. A region is separated from the rest of the cube by an isosurface of 
value $T_{\mathrm{edge}}$, where $T>T_{\mathrm{edge}}$ inside the region and  
 $T<T_{\mathrm{edge}}$ just outside the region. The major and minor axes of the 
region are estimated as the mean values of the second spatial moments in the directions of 
these axes, respectively: 

\begin{equation}
\sigma_{\mathrm{maj,min}}=\sqrt{\frac{\sum_i{{T_i(x_i-\overline{x})^2}}}{\sum_i{T_i}}},
\end{equation}

where $x$ represents the points lying on the major or the minor axis of the region and the sum 
is over the pixels inside the region. The radius, assuming a spherical region, 
is $R=\eta\sqrt{\sigma_{\mathrm{maj}}\thinspace\sigma_{\mathrm{min}}}$, where $\eta=1.91$.

The velocity dispersion estimate is the second moment of the velocity axis weighted by the data values
\begin{equation}
 \sigma_v=\sqrt{\frac{\sum_i ^{\mathrm{region}} T_i(v_i-\overline{v})^2}{\sum_i ^{\mathrm{region}} T_i}}
\end{equation}
where 
\begin{equation}
 \overline{v}=\frac{\sum_i ^{\mathrm{region}}T_i v_i}{\sum_i ^{\mathrm{region}}T_i}
\end{equation}

and the sum is over all the pixels inside the region defined by $T_{\mathrm{edge}}$. 

The flux is estimated as the zeroth moment 
$F=\sum_i{T_i \thinspace\delta v\thinspace\delta x\thinspace \delta y}$, 
where $\delta x$, $\delta y$, and $\delta v$ are the pixel sizes. 

 The value of $T_{\mathrm{edge}}$ adopted here is 
$T_{\mathrm{edge}}=4\times l_{\mathrm{rms}}$. 
However, the method is unbiased with respect to $T_{\mathrm{edge}}$. The rms of 
the datacube is used to set the size of the intervals where 
the method looks for new clouds. We set this parameter as $4\times l_{\mathrm{rms}}$, and 
a minimum pixel area set by the resolution of each observation. \cite{2008ApJ...679.1338R} 
describe three paradigms for measuring the properties from datacubes, the 
one used here is the \textgravedbl bijection\textacutedbl paradigm.

\subsection{Properties of H{\sc{ii}} regions}

\subsubsection{Dust attenuation corrected H$\alpha$ luminosity}

The Antennae have quite heavy and variable attenuation \citep{2005ApJ...635..280B}, 
as expected for a system with widespread ongoing massive star formation.  
This implies that the observed H$\alpha$ does not take fully into account the H{\sc{ii}} region population, 
as the most deeply embedded regions are not observed. We have made the best possible correction for dust extinction and 
believe that the statistical impact of the missing regions is not crucial. 
In order to correct
for this dust attenuation, we used archival \emph{HST} data of The Antennae galaxies with narrow band 
filters in H$\alpha$ F658N, and H$\beta$ F487N. Using the broadband filter F814W (close to I) we  
subtracted the continuum in the H$\alpha$ and H$\beta$ images following the procedure described 
in \cite{2011AJ....141..113G}. Assuming case B recombination 
the intrinsic Balmer line ratio is 
$L_{{\mathrm H\alpha}}/L_{{\mathrm H\beta}}=2.87$ \citep{1989agna.book.....O}, 
and assuming a value of the attenuation curve estimated from \cite{2000ApJ...533..682C} of $K({\mathrm H\alpha}) = 3.33$ and 
$K({\mathrm H\beta}) = 4.60$ we have made an individual correction of
the H$\alpha$ luminosity of each H{\sc{ii}} region pixel by pixel for dust attenuation in the Antennae. 
 The colour excess is $E(B-V)=1.38$ around the NGC 4038 nuclei and $E(B-V)=0.68$ around the NGC 4039 nuclei. 
In the rest of the galaxies, except in the overlap region $E(B-V)\in[0,1]$ while the overlap 
region between the two galaxies is where the maximum colour excess is measured $E(B-V)\in[0,1.84]$. These values are representative, 
and in practice we corrected each pixel individually for extinction. 

The results 
we present here are corrected for this dust attenuation, although the results without this correction 
are qualitatively similar.  

\subsubsection{Derived properties of H{\sc{ii}} regions}

Given the dust-corrected H$\alpha$ luminosity, $L_{{\mathrm H\alpha}}$, the radius of the H{\sc{ii}} region, 
$R_{\mathrm{HII}}$, and the velocity dispersion, $\sigma_{v}$, we can derive a number of physical parameters, e.g.
the electron density, $n_e$, following \cite{2005A&A...431..235R} (equation 6), assuming spherical H{\sc{ii}} regions 
composed of hydrogen with uniform density \citep{1978ppim.book.....S}

\begin{equation}
\frac{L_{{\mathrm H\alpha}}}{\pi {R_{\mathrm{HII}}^{\mathrm{cm}}}^2}=h\nu_{{\mathrm H\alpha}}\alpha_{{\mathrm H\alpha}}^{\mathrm{eff}}(H_0,T)2.46\cdot 10^{17}\cdot n_e^2{R_{\mathrm{HII}}^{\mathrm{cm}}}
\label{eq:density}
\end{equation}

where $\alpha_{{\mathrm H\alpha}}^{\mathrm{eff}}(H_0,T)$ is the effective recombination coefficient of the H$\alpha$ emission
line, $h\nu_{{\mathrm H\alpha}}$ is the energy of an H$\alpha$ photon, and ${R_{\mathrm{HII}}^{\mathrm{cm}}}$ is the radius 
in cm of the H{\sc{ii}} region.  We implicitly assume that the H$\alpha$ emission is due to photoionization. However, 
observations have suggested the presence of shocks in the Antenna 
\citep{2005A&A...433L..17H,2012A&A...538L...9H,2012ApJ...750..136W}. 
Diagnostics to estimate the amount 
of photoionization versus ionization by shocks have been made for a few galaxies yielding the result that 
a rather small fraction of the ionization is due to non-photoionization processes \citep{2004AJ....127.1405C}, about $3\%-4\%$.  
However the diagnostics have a degree of degeneracy, so that in an extreme case the fraction of ionization by shocks can reach
$33\%$ \citep{2011ApJ...731...45H}, although these authors give canonical values of less than 15\% in their study. 
It is necessary to specify here that equation \ref{eq:density} is valid where 
there is not a significant contribution from ionization by shocks.

Using equation \ref{eq:density}, the ionized gas mass ($M_{\mathrm{HII}}$) is given by:

\begin{equation}
M_{\mathrm{HII}}(\mathrm{M_{\odot}})  =  \frac{4}{3}\pi\thinspace R_{\mathrm{HII}}^3 \thinspace n_e \thinspace m_p 
= 1.57\times 10^{-17} \sqrt{L_{{\mathrm H\alpha}}\times R_{\mathrm{HII}}^3}
\label{eq2}
\end{equation}

where $m_p=1.67\times 10^{-27}\mathrm{kg}$ is the proton mass, $L_{{\mathrm H\alpha}}$
is in erg/s, and $R_{\mathrm{HII}}$ is in pc. 

 We have subtracted the instrumental, natural and
thermal widths in quadrature from the observed velocity dispersion. We estimate a  
natural width of $3\thinspace\mathrm{km/s}$ using the data from \cite{1999A&AS..135..359C} 
for the seven fine structure components of H$\alpha$ for an H{\sc{ii}} region with $N_\mathrm{e}\sim10^2$ and 
$T\sim10^4 \thinspace \mathrm{K}$. 
We consider a thermal width of an isothermal H{\sc{ii}} region 
with temperature $T=10^4\thinspace\mathrm{K}$ of $9.1\thinspace\mathrm{km/s}$ 
\citep{1988A&A...198..283O}.

Using the non-thermal
velocity dispersion and the ionized gas mass, $M_{\mathrm{HII}}$,
we can approximate the virial parameter $\alpha_{vir}= 5 \frac{\sigma_{v}^2 R_{\mathrm{HII}}}{GM_{\mathrm{gas}}}$ 
\citep{1992ApJ...395..140B} assuming 
$M_{\mathrm{gas}}=M_{\mathrm{HII}}$. In an H{\sc{ii}} region there is more mass involved than the ionized gas. However, the 
results we will present below comparing properties of H{\sc{ii}} regions and molecular clouds, 
enable us to compare ionized gas with molecular gas mass and conclude that the 
fraction of ionized gas over total gas mass must be nearly constant. 
Under these conditions the true values of the virial parameter for the H{\sc{ii}} regions must be a nearly constant 
fraction of those estimated here. In practice our estimate of the virial parameter is uncertain 
by an essentially constant factor due to the non uniform density distributions and deviations from sphericity 
in clumps \citep{1992ApJ...395..140B}.

We also derive the star formation rates (SFR) for each H{\sc{ii}} region 
using the dust-corrected H$\alpha$ luminosity and the calibration in \cite{2009ApJ...703.1672K}. 
 \cite{2009ApJ...699.1982B} found $\mathrm{SFR}\in[0.2,2.0]\mathrm{M_{\odot}/\mathrm{yr}}$ 
in the Antennae for the brightest 
star forming regions compatible with our SFR estimated for the brightest H{\sc{ii}} regions 
$[2.8,0.8,0.68,0.3,\ldots]\mathrm{M_{\odot}/\mathrm{yr}}$ (see Table \ref{table_hii}).

\subsection{Properties of molecular clouds}
As in  \cite{2012ApJ...750..136W},
the molecular gas mass was estimated from the CO luminosity
following \cite{1992ApJ...398L..29S}:

\begin{equation}
\label{nada}
\begin{split}
 \frac{L_{CO}}{\mathrm{K}\thinspace \mathrm{km}\thinspace \mathrm{s}^{-1}\thinspace \mathrm{pc}^2} & =
 3.25\thinspace \times 10^7 \left(\frac{\nu_{rest}}{\mathrm{GHz}}\right)^{-2}(1+z)^{-1} \\
 &\times \left(\frac{D_L}{\mathrm{Mpc}}\right)^2 \left(\frac{F^{\prime} _{CO}}{\mathrm{Jy}\thinspace \mathrm{km}\thinspace \mathrm{s}^{-1}}\right)
\end{split}
\end{equation}

where $\nu_{\mathrm{rest}}$ is the rest frequency of the line ($345.796\thinspace \mathrm{GHz}$ in the case of CO(3-2)), $D_{\mathrm{L}}$ is the
luminosity distance, assumed $22\thinspace\mathrm{Mpc}$ along the paper, and $F_{\mathrm{CO}}$ is the velocity-integrated flux measured
in the datacube. Then, using the relation between CO luminosity and molecular gas mass,
$M_{\mathrm{H_2}}=\alpha_{\mathrm{CO}}\thinspace L_{\mathrm{CO}}$ (where $\alpha_{\mathrm{CO}}=2m_{H}X_{\mathrm{CO}}$), we obtain the molecular gas mass, $M_{\mathrm{H_2}}$, assuming
 $\alpha_{\mathrm{CO}}=4.8\thinspace M_{\odot} (\mathrm{K}\thinspace \mathrm{km}\thinspace \mathrm{s}^{-1}\thinspace \mathrm{pc}^2)^{-1}$,
the same value as in \cite{2012ApJ...750..136W} and in \cite{2012ApJ...745...65U}. There is a conflict for the value 
of $\alpha_{\mathrm{CO}}$ in the Antennae, some works claim that the value is a few times smaller than the Galactic value of 
$\alpha_{\mathrm{CO}}=4.5$ \citep{2003ApJ...588..243Z,2013ARA&A..51..207B} while other studies claim that the 
value of $\alpha_{\mathrm{CO}}$ in the Antennae is close to the Galactic value \citep{2003ApJ...599.1049W,2007A&A...466..467S}. 
We have chosen the value close to the 
Galactic value for consistency with \cite{2012ApJ...750..136W}, \cite{2012ApJ...745...65U}, and 
\cite{2010ApJ...723..492R}. We have corrected $\alpha_{\mathrm{CO}}$ by a factor of 1.8 since we are using CO(3-2) intensities, 
and from \cite{2012ApJ...745...65U} we know that the average ratio $\frac{I_{\mathrm{CO}}(1-0)}{I_{\mathrm{CO}}(3-2)}\simeq1.8$ in the 
Antennae galaxies. Furthermore, we have applied a factor of 1.36 in order to take into account helium and heavier 
elements in the molecular cloud masses \citep{1973asqu.book.....A}.

We then went on to estimate the density assuming spherical clouds, $\rho_{\mathrm{mol}}$, and the virial parameter,
$\alpha_{vir}= 5 \frac{\sigma_{v}^2 R}{GM_{\mathrm{gas}}}$, assuming $M_{\mathrm{gas}}=M_{\mathrm{mol}}$  
for the molecular clouds.

\subsection{Uncertainties}

We have followed the procedure (the bootstrapping method) explained in \cite{2006PASP..118..590R}
to derive the
errors in the measured properties. This method consists
of randomly sampling each region allowing repeated values. The error is estimated as the standard deviation of
the new derived properties scaled up by the square root of the number of pixels in one resolution element in
the H$\alpha$ data, and in one beam in the ALMA data.  This uncertainty does not include the intrinsic error 
of the flux in the data cubes.

We show the values of these computed physical parameters for the 303 H{\sc{ii}} regions in Table \ref{table_hii}1,
and the 142 molecular clouds detected in Table \ref{table_mol}2. However, as we are resolution limited at the 
small radius end of 
the distribution of H{\sc{ii}} regions and molecular clouds we removed regions with
an uncertainty in size greater than 20\% except in the comparison of our GMCs with those from 
\cite{2012ApJ...745...65U} (Fig. \ref{fig_gmcloc}) since they do not estimate uncertainties. This comparison would not be 
complete if we did not consider all GMCs detected in our data. 
After removing the regions most affected by the noise, 
the number of H{\sc{ii}} regions is 201 and the number of GMCs is 89.

\section{Results}
\begin{figure*}
 \centering
\begin{tabular}{cc}

\epsfig{file=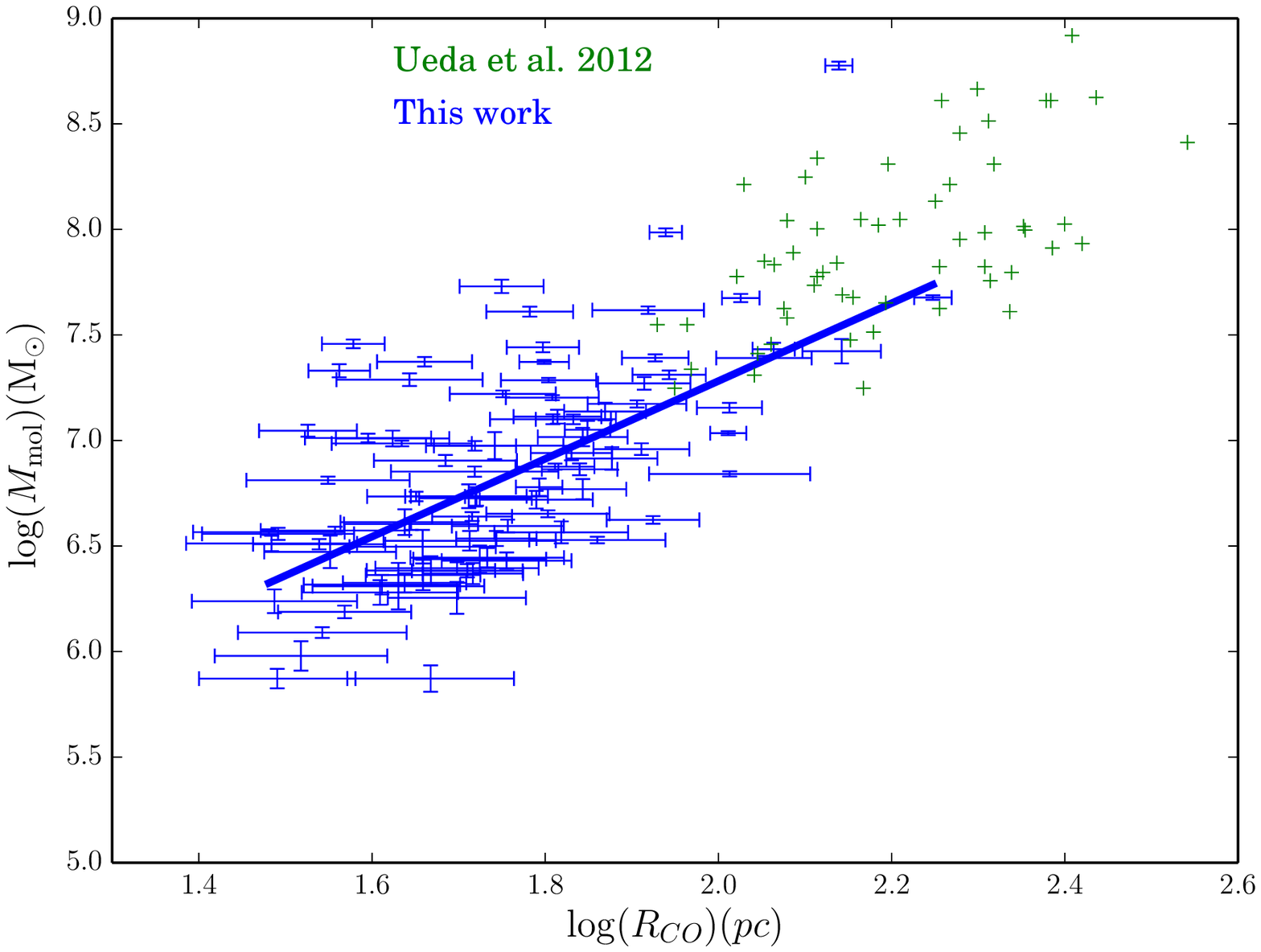,width=0.48\linewidth}

\put(-200,150){\large (a)}
&
\epsfig{file=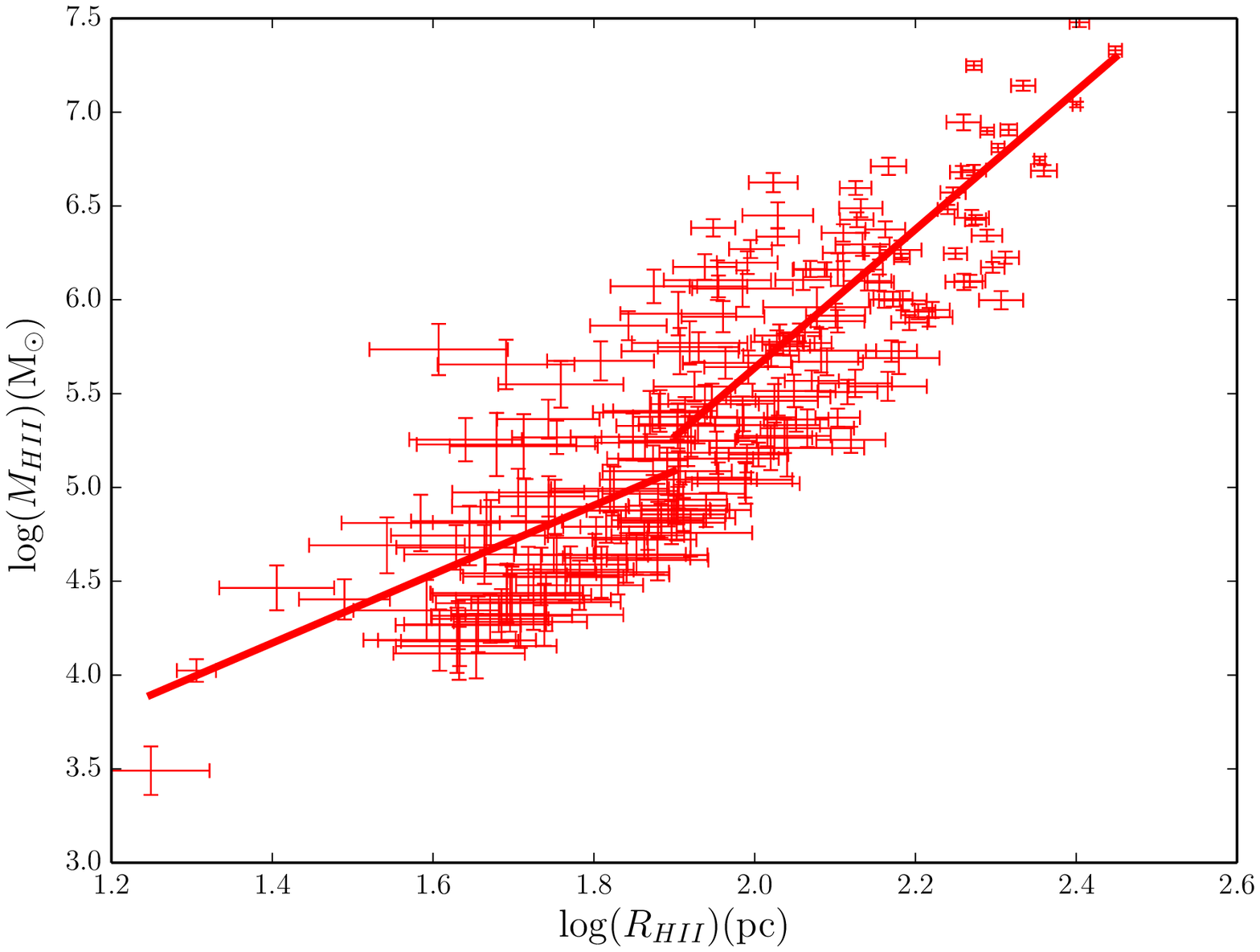,width=0.48\linewidth}

\put(-200,150){\large (b)}

\\

\end{tabular}

\caption{(a) Relation between molecular gas mass, $M_{\mathrm{mol}}$, and  molecular cloud size, $R_{\mathrm{CO}}$, for molecular clouds in the
Antennae using
ALMA data (blue).  The blue line is the linear fit result for molecular clouds using ALMA (equation \ref{polyfit_h2}). 
The same relation is plotted for data of the Antennae from \citet{2012ApJ...745...65U} in green. 
(b) Relation between ionized gas mass, $M_{\mathrm{HII}}$, and  H{\sc{ii}} region size, $R_{\mathrm{HII}}$, for H{\sc{ii}} regions in the
Antennae using
GH$\alpha$FaS data.  The red line is the linear fit result for H{\sc{ii}} regions (equations \ref{polyfit_hii1} and \ref{polyfit_hii2})}
\label{fig_scaling}

\end{figure*}

Fig. \ref{fig_scaling} (a) shows the mass-radius relation for molecular clouds (blue)  compared 
with data of the Antennae from \cite{2012ApJ...745...65U} (green).
Fig. \ref{fig_scaling} (b) shows the same relation for H{\sc{ii}} regions (red).
The scales are different because the sizes of H{\sc{ii}} regions are bigger (as expected since they have been
 \textgravedbl puffed up\textacutedbl during the ionization and heating process) and our estimates of
the H{\sc{ii}} region masses take into account only the ionized gas mass; for the molecular clouds,
we have estimated the molecular gas mass.

 The comparison between our data from ALMA and the data from \cite{2012ApJ...745...65U} using SMA 
shows that they are 
compatible, but our results  have twice better angular resolution and 24 times better sensitivity enabling 
us to resolve larger molecular clouds into smaller ones.
The results of ALMA cover a wider range of masses an sizes compared 
with the results of SMA.

Despite the fact that the H{\sc{ii}} regions have evolved considerably from their original molecular clouds,
H{\sc{ii}} regions and molecular clouds show similar forms for the mass-radius
relation (Fig. \ref{fig_scaling} (a,b)).  Furthermore, assuming that a molecular cloud 
would not be larger than its corresponding evolved H{\sc{ii}} region, we infer that we are not missing 
any flux in the molecular clouds due to the lack of the $u,v$ coverage 
since the largest H{\sc{ii}} region size is $281$pc.

\begin{figure*}

 \begin{adjustbox}{addcode={\begin{minipage}{\width}}{\caption{Molecular gas density, $\rho_{\mathrm{mol}}$, versus 
molecular gas mass of each
molecular cloud, $M_{\mathrm{mol}}$, 
(dark blue). Ionized gas density, 
$\rho_{M_{\mathrm{HII}}}$, 
versus ionized gas mass for each H{\sc{ii}} region (dark red), $M_{\mathrm{HII}}$.
We show the equivalent data for molecular
clouds in the
Milky Way data from \citet{2010ApJ...723..492R} (light blue), and for the H{\sc{ii}} regions of M51  from
\citet{2011AJ....141..113G} (light red).
The blue scales are for the molecular clouds, and the red scales are for the H{\sc{ii}} regions. }\label{fig_dens}\end{minipage}},rotate=90,center}
     \includegraphics[width=1.3\linewidth]{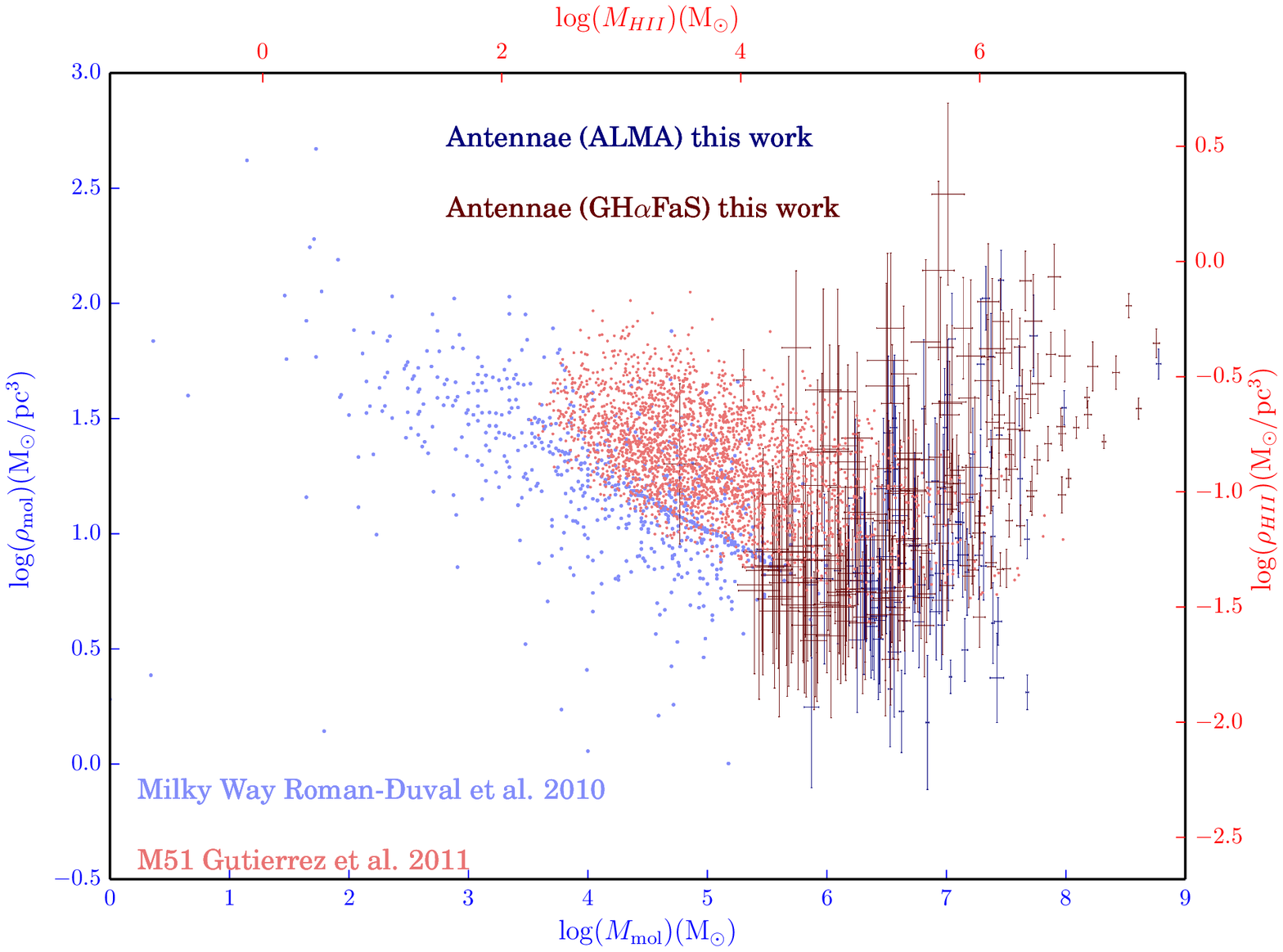}

  \end{adjustbox}

\end{figure*}

\subsection{Molecular clouds results}

In this work, we have found that the exponent $N$ in the mass-radius relation ($M\propto R_{\mathrm{CO}}^N$) 
is significantly higher than 2, in contrast to the results of previous studies for Galactic molecular clouds, 
\cite{1981MNRAS.194..809L} found $N=1.9$, \cite{2010ApJ...723..492R} found $N=2.36$, 
\cite{2010A&A...519L...7L} found $N=1.6$, \cite{2010ApJ...716..433K} found $N=1.33$ and 
extragalactic GMCs where \cite{2014ApJ...784....3C} found $N\in[1.5,2.2]$.

 In Fig. \ref{fig_scaling} (a) we show the linear fit estimated for the molecular clouds 
(blue). The results of the error weighted fit is:

\begin{equation}
\newline
\log(M_{\mathrm{mol}})=(2.4\pm0.4)+(2.62\pm0.10)\log(R_{\mathrm{CO}})
\label{polyfit_h2}
\end{equation}

where $M_{\mathrm{mol}}$ units are $\mathrm{M_{\odot}}$, while $R_{\mathrm{CO}}$ units are $\mathrm{pc}$. We note that 
the exponent is $N=2.62$ which means that the molecular gas surface density increases with the size, 
or with the mass of the cloud, in contrast with 
previous studies where a nearly constant gas surface density is found.

\subsection{H{\sc{ii}} region results}
In order to compare molecular clouds and H{\sc{ii}} region results, we need
to convert the H$\alpha$ luminosity-radius relation to the corresponding ionized gas mass-radius relation.

From equation \ref{eq2} we can infer that if the exponent in the H$\alpha$ luminosity-radius relation
($L_{{\mathrm H\alpha}}\propto R_{\mathrm{HII}}^Q$) is $Q<3$ then the exponent in the ionized gas mass-radius relation
($M_{\mathrm{HII}}\propto R_{\mathrm{HII}}^{N}$) would also be ${N}<3$, and if $Q>3$, then ${N}>3$.

In previous studies of H{\sc{ii}} regions values of $Q<3$ were obtained
\citep{1981MNRAS.195..839T,2011AJ....141..113G}.
However, the most luminous H{\sc{ii}} regions in Arp 270 had $Q=3.8$ in
the $L_{{\mathrm H\alpha}}\propto R_{\mathrm{HII}}^Q$ relation, from results of \cite{2013MNRAS.432..998Z} differing from the 
relation shown above found for essentially non-interacting galaxies.
Instead of using H$\alpha$ luminosity for H{\sc{ii}} regions, we use the ionized gas mass
derived from equation \ref{eq2}.

Here we have performed a double linear fit to the ionized gas mass-radius relation, keeping 
as a fit coefficient the transition between one linear fit and another one.

In Fig. \ref{fig_scaling} (b),
the double linear error weighted fit is shown for H{\sc{ii}} regions (red line). 
The results of the fit are

\begin{equation}
\label{polyfit_hii1}
\begin{split}
\log(M_{HII}) &= 
(1.6\pm0.4)+(1.8\pm0.3)\log(R_{HII}) \thinspace \mathrm{;} \\ 
&\thinspace\thinspace\thinspace\thinspace \mathrm{for}\thinspace \log(R_{HII})\leqslant1.9
\end{split}
\end{equation}

\begin{equation}
\label{polyfit_hii2}
\begin{split}
 \log(M_{HII}) &= 
(-1.7\pm0.5)+(3.7\pm0.2)\log(R_{HII}) \thinspace \mathrm{;} \\ 
&\thinspace\thinspace\thinspace\thinspace \mathrm{for}\thinspace \log(R_{HII})>1.9
\end{split}
\end{equation}

where $M_{\mathrm{HII}}$ and $R_{\mathrm{HII}}$ are in units of $\mathrm{M_{\odot}}$ and $\mathrm{pc}$ respectively. 
The transition at $\log(R_{\mathrm{HII}})=1.92$ corresponds 
to a transition at $\log(M_{\mathrm{HII}})=5.4$ or using equation \ref{eq2}, a transition at $\log L_{{\mathrm H\alpha}}=38.6$. 
For the H{\sc{ii}} regions in the Antennae we see that the less massive regions $\log(M_{\mathrm{HII}})<5.4$ follow the
trend found in previous work ($N<3$ in the $M\propto R_{\mathrm{HII}}^N $ relation). For the more massive 
$\log(M_{\mathrm{HII}})>5.4$ regions
the slope is higher than 3, implying that the density increases with mass, i. e., the star forming 
regions are denser as they grow in mass
(or in size).

\subsection{Density of molecular clouds and H{\sc{ii}} regions}
The change in the mass-size relation is clearly shown in Figure 3 where we show the 
(ionized) gas density-(ionized) gas mass relation for H{\sc{ii}} regions and molecular clouds. We plot two scales 
because, as we explain below, we can make a useful, if somewhat tentative match between the two.
 In Figure 3, the 
comparison of the molecular clouds in the Antennae (dark blue) 
with the molecular clouds in the Milky Way (light blue) shows
a decreasing density regime with increasing mass for the Milky Way molecular clouds, while for those in
the higher mass range of the Antennae molecular clouds, the density
does not decrease with mass. The same pattern occurs for the H{\sc{ii}} regions in the Antennae comparing them with
\cite{2011AJ....141..113G} H{\sc{ii}} regions for M51. 
We considered the use, in Figure 3 of the results of \cite{2014ApJ...784....3C} which we used in section 5,  
on the molecular clouds of M51. However the lower limit to the cloud masses in 
\cite{2014ApJ...784....3C} is only $10^{5}\mathrm{M_{\odot}}$, 
and there are very few clouds with masses below $10^{6}\mathrm{M_{\odot}}$. 
Using these data, we could not
have obtained a useful comparison with the H{\sc{ii}} regions in M51, where the HST 
resolution makes the low mass range available. In order to make an indirect 
comparison we were obliged to use the results of the GMC's in the Milky Way from \cite{2010ApJ...723..492R}.
We have seen in the previous section that the 
transition of mass regimes in the H{\sc{ii}} regions is 
at $\sim10^{5.4} \mathrm{M_{\odot}}$ while for the molecular clouds we do not have enough molecular clouds 
in the lower mass regime in the Antennae, but from Figure 3 it is clear that 
 if we consider the clouds in the Milky Way and the Antennae 
together we find, at least apparently, a transition compatible 
with the transition at $\sim10^{6.5} \mathrm{M_{\odot}}$ of 
the mass function of \cite{2012ApJ...750..136W} and that derived below.  

If we adopt a numerical
factor which brings into registration the scales for the two types of clouds in the 
density-mass relation (Figure 3), we can estimate that
the expansion factor of a GMC which is transformed into an H{\sc{ii}} region is $\sim 2$,
while the ionized gas to total gas fraction is $\sim 0.05$ as predicted in the models by
 \cite{2004A&A...424..877G}. A more reliable estimate of these two parameters is beyond the scope of 
this work. An interesting result here is that when we use the ionized gas mass 
to derive an initial value for the virial parameter, and then go on to find the ionized mass fraction needed to 
convert the virial parameter to values close to unity in the gravitationally bound regime this 
fraction is virtually invariant.


\begin{figure}
 \centering

\epsfig{file=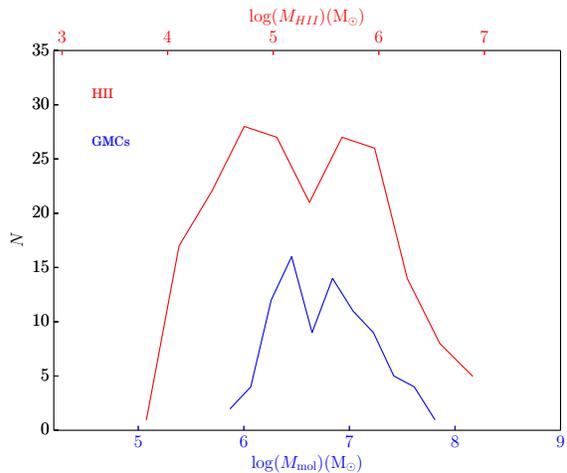,width=0.9\linewidth}

\caption{
Mass function of molecular clouds (blue)
and of H{\sc{ii}} regions (red) in the Antennae. The blue scale  
refers to the 
mass of molecular clouds, $M_{\mathrm{mol}}$, while the red scale 
refers to the ionized gas mass of the H{\sc{ii}} regions, 
$M_{\mathrm{HII}}$.
}
\label{fig_lumfunc}

\end{figure}

We plot the mass
function in Fig. \ref{fig_lumfunc}, which shows the number of regions per mass bin. We find
a break in the molecular gas mass function at $\log(M_{\mathrm{mol}})\sim6.7$, near the value for which 
\cite{2012ApJ...750..136W} found a break in the mass function. The evidence for the presence
of two
populations of molecular clouds is still present in the two populations of H{\sc{ii}} regions with a break at
$\log(M_{\mathrm{HII}})\sim5.4$, which corresponds to a break in the H$\alpha$ luminosity function at
$L_{{\mathrm H\alpha}}\sim38.6\mathrm{dex}$ found
with the GH$\alpha$FaS data. This has also been reported previously in the literature
\citep{1989ApJ...337..761K,2006A&A...459L..13B}. Again, we match the two mass functions
using a constant ionized gas fraction around $0.05$ as we found previously and as was assumed in the models of 
\cite{2004A&A...424..877G}. The constant ionization fraction gives a match to the break change in the 
density-mass relation for molecular clouds at $\log(M_{\mathrm{mol}})\sim6.7$ and H{\sc{ii}} regions at $\log(M_{\mathrm{HII}})\sim5.4$.

\subsection{Location of H{\sc{ii}} regions and GMCs}

\begin{figure}
\centering
\epsfig{file=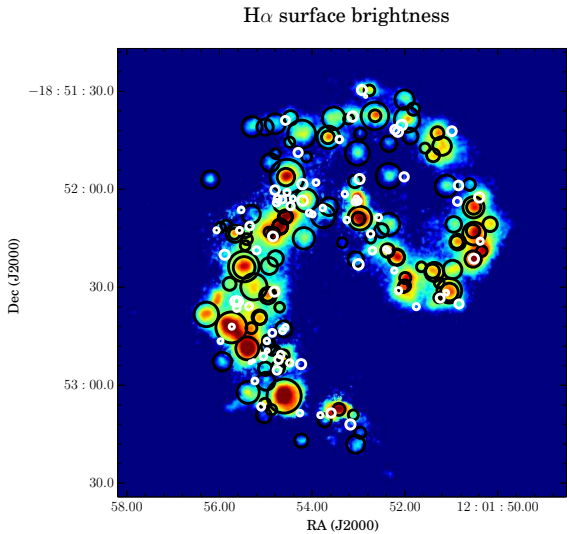,width=0.9\linewidth}
\caption{H{\sc{ii}} regions locations around Antennae galaxies. The intensity colour map is 
the H$\alpha$ surface brightness derived from the GH$\alpha$FaS datacube. 
White circles represent 
the less massive H{\sc{ii}} regions ($\log(M_{\mathrm{HII}})<5.4$) and black circles represent the more 
massive H{\sc{ii}} regions ($\log(M_{\mathrm{HII}})>5.4$)
}
\label{fig_loc}

\end{figure}

In figure \ref{fig_loc} we represent the H$\alpha$ surface brightness map, with the H{\sc{ii}} regions over plotted 
as white and black circles. There are 80 H{\sc{ii}} regions in the less massive regime (white circles in 
Fig. \ref{fig_loc}) while there are 121 H{\sc{ii}} regions in the more massive regime (white circles in 
Fig. \ref{fig_loc}). We do not see an obvious correlation between the two regimes and their locations within 
the galaxies. This result is in contrast with the pattern we found \citep{2013MNRAS.432..998Z} in Arp 270 
between the two regimes where the regime of the high mass regions was concentrated towards the central part 
of the galaxies and the overlap region, while regime of the low mass regions was concentrated in the outer parts of 
the galaxies. We do not see this pattern in the Antennae galaxies almost certainly because they are in a more advanced, 
and therefore morphologically more complex, stage of the merger process.  

\begin{figure}
\centering
\epsfig{file=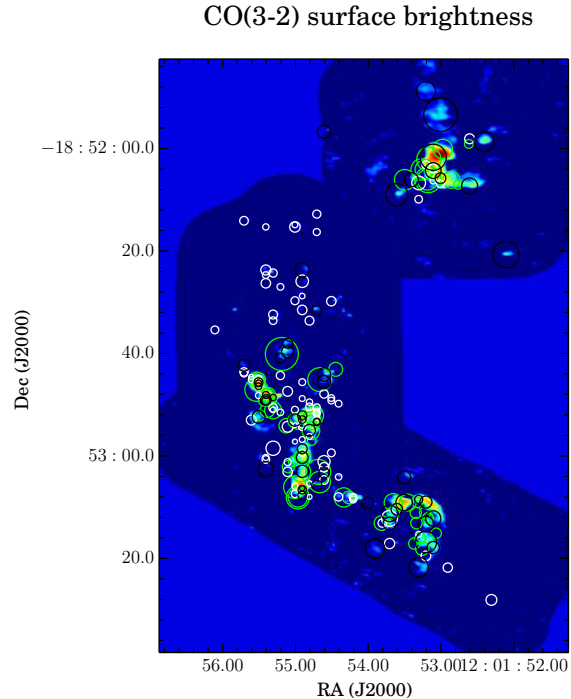,width=0.9\linewidth}
\caption{GMC locations around Antennae galaxies. The intensity colour map is 
the CO(3-2) surface brightness derived from the ALMA datacubes. 
White circles represent 
the less massive GMCs ($\log(M_{\mathrm{mol}})<6.7$), black circles represent the more 
massive GMCs ($\log(M_{\mathrm{mol}})>6.7$), and green circles 
represent the GMCs detected by \citet{2012ApJ...745...65U}
}
\label{fig_gmcloc}

\end{figure}

In Fig. \ref{fig_gmcloc} we show the CO(3-2) surface brightness map with the GMCs represented as black and white 
circles (56 in the more massive regime, and 86 in the less massive regime respectively). In contrast with the previous sections, 
we have used the whole sample of GMCs detected independently of the uncertainty in the size of the region as we 
have explained above. 
The locations of GMCs is not 
correlated with the two types of GMCs, as in the case of H{\sc{ii}} regions. We have plotted, in order to compare, the regions from 
\cite{2012ApJ...745...65U} as green circles. In Fig. \ref{fig_gmcloc}, it is clear that the data 
of this work compared with the data from \cite{2012ApJ...745...65U} are improved because of the better 
spatial coverage, sensitivity, and spatial resolution.

\section{Gravitationally bound regime}

 It is not clear what is the relative importance of the different forces which control the physical behaviour of
molecular clouds and H{\sc{ii}} regions (gravity, magnetic field, external pressure, shear, etc.). 
It also notable that environment may play a specific role in 
the behaviour of star forming regions, and subsequently of star formation. 
Articles with star formation models, and observations, from \cite{1997ApJ...480..235E} and \cite{2001AJ....122.1888A} 
claim that the external pressure is the driver of the star formation in starburst regions, but that 
 the mass and the self-gravity of the star forming regions should also play a role.

 \cite{2012ApJ...750..136W},  claimed that the high mass population of molecular clouds 
is gravitationally bound  based on the relation between virial and molecular gas mass. 
We also note that  \cite{2011AJ....141..113G} showed clearly that the H{\sc{ii}} regions in M51 are pressure 
bounded by the external gas column density, (we note that these are regions with luminosities below $38.6\mathrm{dex}$) while 
\cite{2013MNRAS.432..998Z} claimed that 
for Arp 270 the brightest H{\sc{ii}} regions are different in this respect from those in M51 
based on their scaling relations, and they concluded that the brightest H{\sc{ii}} regions are 
gravitationally dominated while for the fainter H{\sc{ii}} regions, the
dominant force is the external pressure rather than gravity.

\begin{figure*}
\begin{tabular}{cc}

 \centering
\epsfig{file=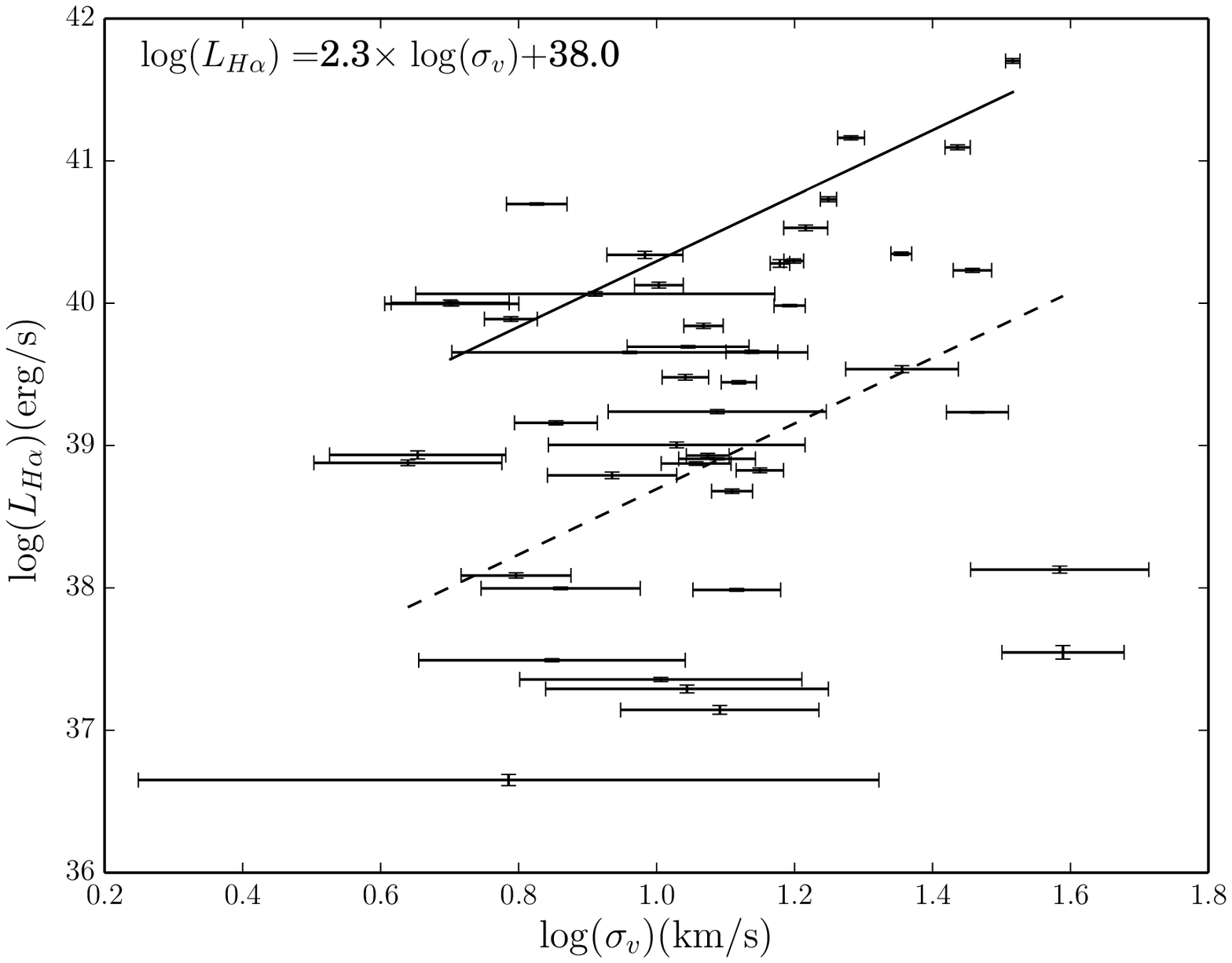,width=0.48\linewidth}

\put(-200,150){\large (a)}
&
\epsfig{file=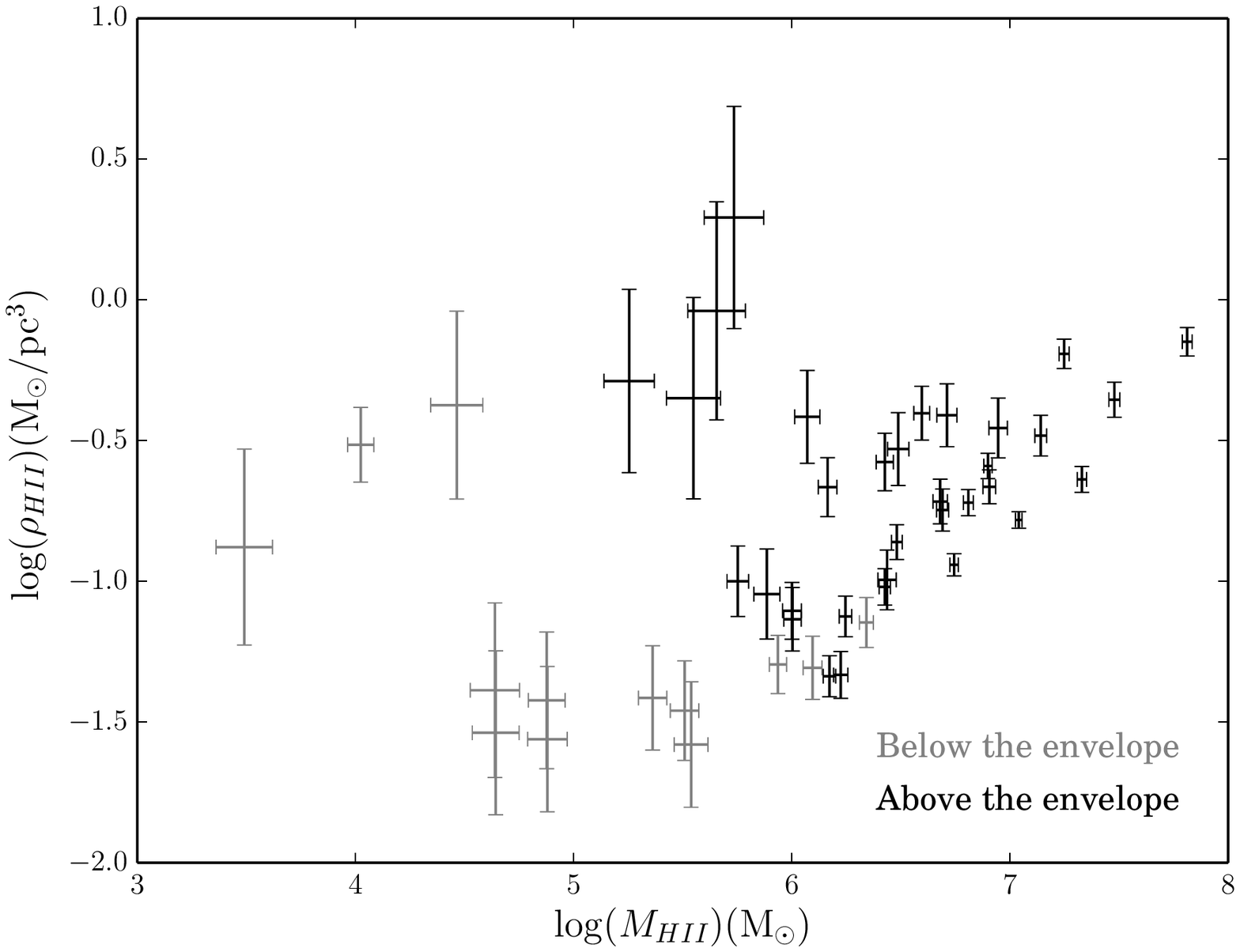,width=0.48\linewidth}

\put(-180,150){\large (b)}
       \\
\end{tabular}

\caption{(a) ${\mathrm H\alpha}$ luminosity, versus velocity dispersion, $\sigma_v$ for the H{\sc{ii}} regions.
The thermal, natural and instrumental widths have been subtracted quadratically. 
The results of the estimates involving the envelope as described here are shown
as a solid line. The dashed line is the same estimate, but shifted by a constant value.
(b) Ionized gas density, $\rho_{\mathrm{HII}}$, versus ionized gas mass, $M_{\mathrm{HII}}$, of the H{\sc{ii}} regions. Regions
in grey represent those which are below the shifted envelope (dashed line) in Fig. \ref{fig_mvir} (a).
Regions in black represent those which are above the shifted envelope (dashed line) in Fig. \ref{fig_mvir} (a).
}
\label{fig_mvir}

\end{figure*}

The tools we have used in the present study to check this hypothesis are the luminosity-velocity 
dispersion ($L_{{\mathrm H\alpha}}-\sigma_v$) relation \citep{2005A&A...431..235R} and 
the virial parameter \citep{1992ApJ...395..140B}. 

The ($L_{{\mathrm H\alpha}}-\sigma_v$) relation for H{\sc{ii}} regions usually presents an envelope through the 
brightest regions. For each luminosity, a range of velocity dispersion values is present. 
The envelope is formed by the regions showing the minimum velocity 
dispersion for given luminosity (see Fig. \ref{fig_mvir} (a)). We 
follow the method of \cite{2005A&A...431..235R} to estimate the envelope.  
They simply selected the H{\sc{ii}} region with the minimum velocity dispersion for each luminosity bin.  
 Furthermore, they suggested that the envelope
in the luminosity-velocity dispersion may be formed by virialized regions i. e. those that 
are gravitationally dominated. 

We have tested this option for the H{\sc{ii}} regions in the Antennae. 
(We  might also have plotted the equivalent envelope in the $M_{\mathrm{mol}}-\sigma_v$
for the molecular clouds, although for these 
we do not have a big enough set of the less massive clouds).

The $L_{{\mathrm H\alpha}}-\sigma_v$ relation is shown in Fig. \ref{fig_mvir} (a) for H{\sc{ii}} regions.
We removed H{\sc{ii}} regions with line widths
less than $8\mathrm{km/s}$, i.e., velocity dispersions less than $4\mathrm{km/s}$, which is the effective
velocity resolution of the
instrument ($8\mathrm{km/s}$ for GH$\alpha$FaS).

We have assumed, for simplicity, that the regions lying on the envelope follow a linear relation in the log-log plane

\begin{equation}
\newline
\log(L_{{\mathrm H\alpha}\thinspace \mathrm{env}})=\beta_1\times \thinspace \log(\sigma_{v})+\beta_0.
 \label{env_fit}
\end{equation}

The result is $\beta_1=2.3$, $\beta_0=38.0$,  accounting just for 
the H{\sc{ii}} regions on the envelope, those with the minimum velocity dispersion 
per luminosity bin. The fit plotted is shown in Figure 
\ref{fig_mvir} (a) as a continuous line. 
 We want to separate the regions that are close to the envelope from those which are not. Thus, 
for a given ($L_{{\mathrm H\alpha}},\sigma_{v}$) values of an H{\sc{ii}} region, we consider that region is 
close to the envelope if $L_{{\mathrm H\alpha}}>\frac{L_{{\mathrm H\alpha}\thinspace \mathrm{env}}}{40}$, where 
$L_{{\mathrm H\alpha}\thinspace \mathrm{env}}$ is derived from equation \ref{env_fit} using the velocity dispersion of 
the region. This criterion is represented as a dashed line in Fig. \ref{fig_mvir} (a).
The relation between the gas mass and the gas density is plotted in \ref{fig_mvir} (b). The regions
in this plot are separated as previously explained, grey symbols represent regions below the shifted
fit of the envelope, while black symbols represent regions above the shifted envelope. It is
clear in Fig. \ref{fig_mvir} (b) that regions above the shifted envelope belong to the more massive regime
where the gas density increases with mass. Regions below the shifted envelope belong to the less massive regime
where gas density decreases with mass.

In addition, the H$\alpha$ luminosity of the regions lying on the unshifted envelope, or as explained above,
the regions in the higher luminosity (or more massive) regime , is correlated super-linearly
with the velocity dispersion. Thus, the SFR derived from the H$\alpha$ luminosity
is correlated with the velocity dispersion for the more massive regions. 
 Although H$\alpha$ is an imperfect measure of the SFR, 
our values are carefully dust corrected and we believe that further refinement 
would not change any of the conclusions.

The virial parameter, $\alpha_{vir}= 5 \frac{\sigma_{v}^2 R}{GM_{\mathrm{gas}}}$, can be used to 
make a test between a regime dominated by external pressure and a gravitationally 
dominated regime. 
$\alpha_{vir}\gg1$ represent regions confined by external pressure where the self-gravity does not dominate.
However,
$\alpha_{vir}\backsimeq1$ represent regions where the velocity dispersion of the cloud can be accounted for
by gravity. In this case, the gravity is the driver of the velocity dispersion and it is the dominant force
\citep{1992ApJ...395..140B}.

\begin{figure}
\centering
\epsfig{file=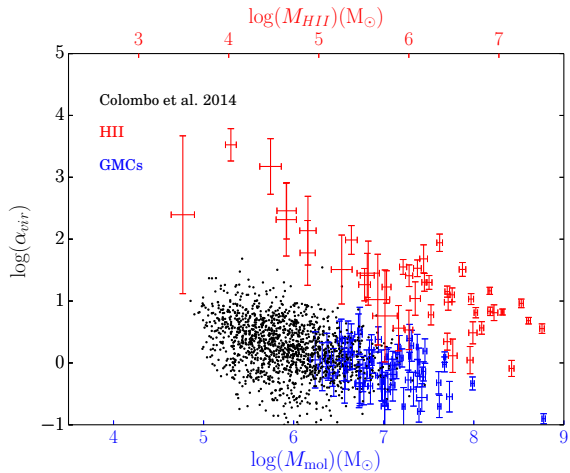,width=0.9\linewidth}
\caption{Virial parameter ($\alpha_{vir}= 5 \frac{\sigma_{v}^2 R}{GM_{\mathrm{gas}}}$)
versus the ionized gas mass, $M_{\mathrm{HII}}$, for the H{\sc{ii}} regions (red) and versus molecular 
gas, $M_{\mathrm{mol}}$, for the molecular clouds in this work (blue) and for M51 
(black) from \citet{2014ApJ...784....3C}. 
In the estimation of the virial parameter, 
$R=R_{\mathrm{HII}}$ for H{\sc{ii}} regions while 
$R=R_{\mathrm{CO}}$ for the molecular clouds, and $M_{\mathrm{gas}}=M_{\mathrm{HII}}$ for H{\sc{ii}} regions while 
$M_{\mathrm{gas}}=M_{\mathrm{mol}}$ for molecular clouds. 
}
\label{fig_virpar}

\end{figure}

We plot the virial parameter versus the gas mass in Fig. \ref{fig_virpar}.
The reason why the virial parameter reaches values smaller than $1$
in the molecular clouds
is due to the effects of nonuniformity in density and deviations
from sphericity.

For the  H{\sc{ii}} regions the virial
parameter value for the less massive regions is greater than unity as expected if 
the external pressure is the dominant force. In the more massive H{\sc{ii}} regions the virial parameter 
decreases, as expected if gravity becomes dominant as the mass increases, supporting the
hypothesis that the difference between regimes is due to different dominant forces. In the 
case of GMCs we do not detect clouds in the less massive regimes. We have added the 
GMCs detected in \cite{2014ApJ...784....3C} for M51 since they estimated the virial parameter. 
Fig. \ref{fig_virpar} shows that the virial parameter decreases with molecular gas mass 
combining GMCs from both data sets, implying 
that external pressure becomes more important in the less massive regions. However, in the 
more massive regions the external pressure looses its importance in favour of self-gravity. 
Although we have combined data from two different systems, the result while not 
conclusive is certainly suggestive.

\begin{figure}
\centering
\epsfig{file=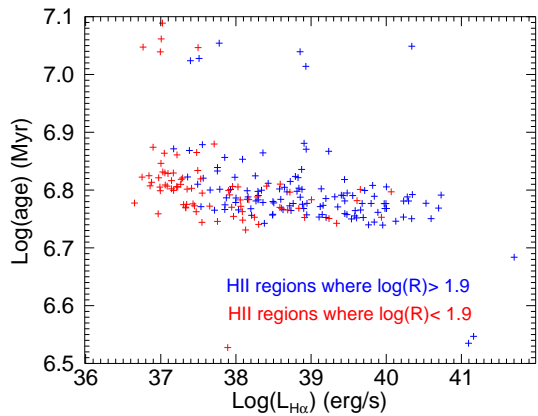,width=0.9\linewidth}
\caption{H{\sc{ii}} region ages derived using the equivalent width and Starburst99 model \citep{1999ApJS..123....3L}
 versus H$\alpha$ luminosity ($L_{{\mathrm H\alpha}}$). 
}
\label{fig_eqw}

\end{figure}

\subsection{Evolutionary processes}
Evolution may play a role in the scaling relations of H{\sc{ii}} regions. However, in \cite{2013MNRAS.432..998Z} it was shown 
that the two populations of H{\sc{ii}} regions in Arp 270 could not be explained, using the H$\alpha$ 
equivalent width, $W(\mathrm H\alpha)$, as an effect of age.  We can assume a direct relation between $W(\mathrm H\alpha)$ and the age of the region 
\citep{1999ApJS..123....3L}. The younger the region, the larger is its equivalent width.

We have used the H$\alpha$ map and the continuum map to derive the H$\alpha$ 
equivalent width. Fig. \ref{fig_eqw} shows the H{\sc{ii}} region ages assuming solar metallicity. 
We can see that the age ranges for two populations of H{\sc{ii}} regions are 
the same. The H{\sc{ii}} regions are around 
$6 \mathrm{Myr}$ old. 
Thus, H{\sc{ii}} region ages do not play an important role in the scaling relations of the two regimes in the Antennae.

\section{Conclusion}

The scaling relations for the most massive clouds
are different from those found for the Milky Way or for non-merging galaxies, where the upper mass
limits for these types of objects are lower.

Previous studies show that gas density decreases with mass \cite{2010ApJ...723..492R,2010ApJ...716..433K,2011AJ....141..113G}.
We showed some of these results in Figure 3, where the gas density of the
less massive molecular clouds of the Milky Way and
the less massive H{\sc{ii}} regions of M51 decrease with mass.

Furthermore, the properties of the more massive molecular clouds in
the Antennae galaxies using archive ALMA data are different from those of the less massive ones. 
The gas density does not decrease with mass. We have been able
to see both regimes in the ionized gas density of the H{\sc{ii}} regions, one where the ionized gas density increases with ionized mass
(for $M_{\mathrm{HII}}>10^{5.4}\mathrm{M_{\odot}}$), and another one where the ionized gas density decreases with ionized gas mass
(for $M_{\mathrm{HII}}<10^{5.4}\mathrm{M_{\odot}}$). The transition is equivalent to the transition found in the mass function 
of molecular clouds at $M_{\mathrm{mol}}=10^{6.7}\mathrm{M_{\odot}}$ in this work and in \cite{2012ApJ...750..136W} if we consider 
a constant ionization fraction of $0.05$, consistent with results found in this work and with models of 
\cite{2004A&A...424..877G}.

In the more massive regime, the more massive the regions are the denser they become.
As the SFR does not appear to be linear with the gas density \citep{1959ApJ...129..243S,1998ApJ...498..541K}, 
the SFR may well be preferentially triggered in these, the most massive clouds.

We have found that there are H{\sc{ii}} regions where the
H$\alpha$ luminosity is correlated with the velocity dispersion. This means that the SFR
derived
from H$\alpha$ luminosity varies approximately with the square of the velocity dispersion, implying 
that the regions on the envelope of the luminosity-velocity dispersion relation are probably
gravitationally dominated, and this is suggested also by the present study. 
Moreover, the H{\sc{ii}} regions on the envelope are those where the gas density increases with mass, 
while for the H{\sc{ii}} regions far from the envelope the gas density decreases with mass.

The virial parameter decreases with increasing gas mass implying that gravity rather than external pressure 
is more dominant for
the more massive H{\sc{ii}} regions and quite probably for the GMCs, 
but the situation is reversed in the 
less massive regions, 
which offers an explanation for why the most massive clouds, be they molecular clouds or H{\sc{ii}} regions, 
tend to be denser with increasing mass.

\section*{Acknowledgements}
We thank Johan H. Knapen for allowing us to use observing time with GHaFaS.
Based on observations made with the William Herschel Telescope operated on the island of La Palma
by the Isaac Newton Group of Telescopes in the Spanish Observatorio del Roque de los
Muchachos of the
Instituto de Astrofísica de Canarias. This research has been supported by the Spanish
Ministry of Economy and Competitiveness (MINECO) under the grants
AYA2007-67625-CO2-01, AYA2009-12903 and  AYA2012-39408-C02-02.
JEB acknowledges financial support to the DAGAL network from the People Programme
(Marie Curie Actions) of the European Union's Seventh Framework Programme FP7/2007-2013/
under REA grant agreement number PITN-GA-2011-289313.

This research made use of \textsc{astropy}, a community-developed core textsc{python} package for 
astronomy \citep{2013A&A...558A..33A},
\textsc{aplpy}, an open-source plotting package for textsc{python} hosted at http://aplpy.github.com, and \textsc{astrodendro},
a textsc{python} package to compute dendrograms of Astronomical data (http://www.dendrograms.org/).
The Atacama Large Millimeter/submillimeter Array (ALMA), an international astronomy facility,
is a partnership of Europe, North America and East Asia in cooperation with the Republic of Chile.
This paper makes use of the following ALMA data:
\newline
ADS/JAO.ALMA\#2011.0.00003.SV  ADS/JAO.ALMA\#2011.0.00876.S

Based on observations made with the NASA/ESA Hubble Space Telescope, obtained from the data archive at 
the Space Telescope Science Institute. STScI is operated by the Association of Universities 
for Research in Astronomy, Inc. under NASA contract NAS 5-26555.

We thank the anonymous referee, whose comments have led to important improvements on the original version of the paper.

\bibliographystyle{mn2e}


\appendix
\section{Properties of the H{\sc{ii}} regions and the GMC's }
\begin{table*}
\centering
 \begin{minipage}{140mm}

  \caption{Physical properties of the brightest H{\sc{ii}} regions derived as described in section $\S3$. 
The whole table 
is available as a machine readable table in the electronic version of the paper and through CDS. }
  \begin{tabular}{@{}ccccccccccc@{}}

  \hline
 N  &RA&Dec & $\log(L_{{\mathrm H\alpha}}) $ & $\mathrm{SFR}$& $R_{\mathrm{HII}}$ & $\log(M_{\mathrm{HII}})$ & $\rho_{\mathrm{HII}}$  &$\sigma_{v}$& $\alpha_{vir}$ \\
  &     hh:mm:ss         &     $^{\circ}$ $\mathrm{\prime}$ $\mathrm{\prime\prime}$ &  $\mathrm{erg/s}$ & $\mathrm{M_{\odot}/\mathrm{yr}}$  &   pc  &$\mathrm{M_{\odot}}$ & $\mathrm{M_{\odot}/pc^3}$ &         $\, \mathrm{km/s}$    &                                   \\



  \hline
\\
    1 & 12: 1:55.6 & -18:52:52.0 & $ 41.7\pm0.03 $ & 2.8 & $ 280\pm5 $ & $ 7.81\pm0.05 $ & $ 0.71\pm0.03 $ & $ 32.8\pm0.7 $ & $ 5.4\pm0.6 $\\   
    2 & 12: 1:53.9 & -18:52:42.7 & $ 41.16\pm0.03 $ & 0.8 & $ 253\pm6 $ & $ 7.48\pm0.05 $ & $ 0.44\pm0.02 $ & $ 19.1\pm0.7 $ & $ 3.6\pm0.5 $\\   
    3 & 12: 1:50.9 & -18:52:24.7 & $ 41.1\pm0.03 $ & 0.68 & $ 187\pm4 $ & $ 7.25\pm0.05 $ & $ 0.64\pm0.03 $ &  $ 27\pm1 $ &   $ 9\pm1 $\\   
    4 & 12: 1:51.2 & -18:52:33.1 & $ 40.73\pm0.04 $ & 0.3 & $ 281\pm5 $ & $ 7.33\pm0.04 $ & $ 0.23\pm0.01 $ & $ 17.7\pm0.4 $ & $ 4.8\pm0.5 $\\   
    5 & 12: 1:53.2 & -18:53: 4.3 & $ 40.35\pm0.02 $ & 0.12 & $ 195\pm3 $ & $ 6.9\pm0.04 $ & $ 0.26\pm0.01 $ & $ 22.6\pm0.7 $ &  $ 15\pm2 $\\   
    6 & 12: 1:55.3 & -18:52:52.8 & $ 40.3\pm0.03 $ & 0.11 & $ 251\pm2 $ & $ 7.04\pm0.03 $ & $ 0.165\pm0.005 $ & $ 15.8\pm0.5 $ & $ 6.6\pm0.6 $\\   
    7 & 12: 1:55.3 & -18:52:52.4 & $ 39.98\pm0.02 $ & 0.053 & $ 187\pm6 $ & $ 6.69\pm0.06 $ & $ 0.18\pm0.01 $ & $ 15.6\pm0.7 $ &  $ 11\pm2 $\\   
    8 & 12: 1:51.4 & -18:52:25.1 & $ 39.84\pm0.04 $ & 0.038 & $ 226\pm3 $ & $ 6.74\pm0.04 $ & $ 0.114\pm0.004 $ & $ 11.7\pm0.7 $ &   $ 6\pm1 $\\   
    9 & 12: 1:54.1 & -18:52: 2.7 & $ 39.89\pm0.03 $ & 0.043 & $ 134\pm6 $ & $ 6.43\pm0.08 $ & $ 0.27\pm0.02 $ & $ 6.1\pm0.5 $ & $ 2.2\pm0.6 $\\   
   10 & 12: 1:51.0 & -18:52:30.8 & $ 40.13\pm0.04 $ & 0.074 & $ 201\pm3 $ & $ 6.81\pm0.05 $ & $ 0.19\pm0.009 $ & $ 10.1\pm0.7 $ & $ 3.7\pm0.7 $\\   

  \end{tabular}
    \end{minipage}
 \label{table_hii}
\end{table*}

\begin{table*}
\centering
 \begin{minipage}{140mm}
\caption{Physical properties of the brightest molecular clouds derived as described in section $\S3$. 
The whole table is available as a machine readable table in the electronic version of the paper and through CDS. }

  \begin{tabular}{@{}cccccccccc@{}}
  \hline
 N &RA&Dec  &$\log(L_{\mathrm{CO}})$ & $R_{\mathrm{CO}}$ & $\log(M_{\mathrm{mol}})$ & $\rho_{\mathrm{mol}}$  &$\sigma_{v}$& $\alpha_{vir}$ \\
  &     (hh:mm:ss)         &     ($^{\circ}$ $\mathrm{\prime}$ $\mathrm{\prime\prime}$)   &$\mathrm{K\thinspace km\thinspace s^{-1}\thinspace pc^2}$&   (pc)  &$\mathrm{M_{\odot}}$ & $\mathrm{M_{\odot}/pc^3}$ &         $\, \mathrm{km/s}$                         &                                   \\
  \hline
\\
    1 & 12: 1:53.1 & -18:52: 1.6 & $ 7.71\pm0.04 $ & $ 138\pm4 $ & $ 8.78\pm0.04 $ &  $ 55\pm7 $ &  $ 22\pm1 $ & $ 0.13\pm0.02 $\\   
    2 & 12: 1:53.5 & -18:53: 9.0 & $ 6.92\pm0.04 $ &  $ 87\pm3 $ & $ 7.99\pm0.04 $ &  $ 35\pm5 $ &  $ 21\pm1 $ & $ 0.5\pm0.1 $\\   
    3 & 12: 1:55.4 & -18:52:49.0 & $ 6.66\pm0.06 $ &  $ 56\pm5 $ & $ 7.73\pm0.06 $ & $ 70\pm30 $ &  $ 15\pm3 $ & $ 0.3\pm0.1 $\\   
    4 & 12: 1:53.0 & -18:51:53.3 & $ 6.61\pm0.02 $ & $ 177\pm8 $ & $ 7.68\pm0.02 $ & $ 2.1\pm0.3 $ & $ 16.9\pm0.9 $ & $ 1.2\pm0.2 $\\   
    5 & 12: 1:52.4 & -18:51:58.6 & $ 6.61\pm0.04 $ & $ 106\pm5 $ & $ 7.68\pm0.04 $ &   $ 9\pm2 $ &  $ 19\pm2 $ & $ 0.9\pm0.2 $\\   
    6 & 12: 1:53.1 & -18:52: 4.2 & $ 6.55\pm0.04 $ & $ 80\pm10 $ & $ 7.62\pm0.04 $ &  $ 17\pm7 $ & $ 9.2\pm0.9 $ & $ 0.2\pm0.07 $\\   
    7 & 12: 1:53.6 & -18:53:10.3 & $ 6.54\pm0.05 $ &  $ 61\pm6 $ & $ 7.61\pm0.05 $ & $ 40\pm20 $ &  $ 17\pm2 $ & $ 0.5\pm0.2 $\\   
    8 & 12: 1:54.7 & -18:52:53.3 & $ 6.39\pm0.04 $ &  $ 38\pm3 $ & $ 7.46\pm0.04 $ & $ 130\pm30 $ &  $ 32\pm4 $ & $ 1.6\pm0.6 $\\   
    9 & 12: 1:54.8 & -18:52:54.9 & $ 6.37\pm0.05 $ &  $ 63\pm5 $ & $ 7.44\pm0.05 $ &  $ 27\pm8 $ &  $ 17\pm1 $ & $ 0.7\pm0.2 $\\   
   10 & 12: 1:53.6 & -18:52: 8.9 & $ 6.36\pm0.06 $ & $ 116\pm6 $ & $ 7.43\pm0.06 $ & $ 4.2\pm0.9 $ &  $ 11\pm1 $ & $ 0.6\pm0.2 $\\   

  \end{tabular}
    \end{minipage}
 \label{table_mol}
\end{table*}

\end{document}